\documentclass[aps,onecolumn,superscriptaddress,nofootinbib]{revtex4-2}
\usepackage{bm}
\usepackage{bbm}
\usepackage{epsfig,graphics,graphicx}
\usepackage{amsmath,amssymb,amsfonts}
\usepackage{color}
\usepackage{epstopdf}
\usepackage{slashed}
\usepackage[
colorlinks=true,
linkcolor=black,
breaklinks=true,
urlcolor=blue,
citecolor=fuchsia]{hyperref}

\usepackage[font=footnotesize, labelfont=bf]{caption,subcaption}
\usepackage{booktabs}
\usepackage{dcolumn}
\usepackage{makecell}
\usepackage{multirow}
\newcolumntype{d}{D{.}{.}{5}}

\definecolor{blueviolet}{rgb}{0.541, 0.169, 0.886}
\definecolor{fuchsia}{rgb}{1.0, 0, 1.0}
\newcommand{\be}{\begin{equation}}
	\newcommand{\ee}{\end{equation}}
\newcommand{\ba}{\begin{eqnarray}}
	\newcommand{\ea}{\end{eqnarray}}

\begin{document}
\title{Precision calculation of the recoil--finite-size correction for the hyperfine splitting in
  muonic and electronic hydrogen}
\author{Aldo Antognini}
\affiliation{Laboratory for Particle Physics, Paul Scherrer Institute, 5232 Villigen-PSI,
Switzerland}
\affiliation{Institute for Particle Physics and Astrophysics, ETH, 8093 Zurich, Switzerland}
\author{Yong-Hui Lin}
\affiliation{Helmholtz Institut f\"ur Strahlen- und Kernphysik and Bethe Center
   for Theoretical Physics, Universit\"at Bonn, D-53115 Bonn, Germany}
\author{Ulf-G. Mei{\ss}ner}
\affiliation{Helmholtz Institut f\"ur Strahlen- und Kernphysik and Bethe Center
   for Theoretical Physics, Universit\"at Bonn, D-53115 Bonn, Germany}
\affiliation{Institute for Advanced Simulation and Institut f{\"u}r Kernphysik,
            Forschungszentrum J{\"u}lich, D-52425 J{\"u}lich, Germany}
\affiliation{Tbilisi State University, 0186 Tbilisi, Georgia}
\date{\today}
%
\begin{abstract}
  We present a high-precision calculation of the recoil--finite-size correction to the hyperfine
  splitting (HFS) in muonic and electronic hydrogen based on nucleon electromagnetic form
  factors obtained from dispersion theory. This will help guide the upcoming searches of the HFS transition in muonic hydrogen, and will allow a precise determination of the polarizability and Zemach radius contributions when this transition is found.
\end{abstract}
\maketitle

\section{Introduction}

Laser spectroscopy of muonic hydrogen ($\mu$p), an atom formed by a negatively charged
muon and a proton,
represents an excellent pathway to investigate low-energy properties of the proton.
The exquisite sensitivity of the muonic hydrogen energy levels to the proton structure rests
on the large muon mass, 207 times larger than the electron mass, that leads to a $10^7$ times
larger overlap between the atomic wavefunction and the proton compared to regular (electronic)
hydrogen, abbreviated as H in what follows.

The measurement of the 2S-2P energy splitting by the CREMA collaboration  with $1\times 10^{-5}$
relative accuracy~\cite{Pohl:2010zza,Antognini:2013txn} and its comparison with the 
corresponding theoretical prediction (we use here the updated theory of Ref.~\cite{Antognini:2022xoo})
\begin{equation}
 E^\mathrm{th}_\mathrm{2P-2S}=206.03470(3)- 5.2275(10)r_p^2 - \Delta E^{2\gamma}_\text{2S-2P} \quad \mbox{[meV]}
\end{equation}
can be used either to extract the proton charge radius $r_p$ with unprecedented accuracy when assuming
the two-photon exchange contribution $\Delta E^{2\gamma}_\text{2S-2P} $ from theory (and measured data
from electron-proton scattering), or to extract $\Delta E^{2\gamma} $ when assuming a proton
charge radius from H or electron-proton scattering.
Using the best data-driven evaluation of the $2\gamma$-exchange
$\Delta E^{2 \gamma }_\text{2S-2P} = -33(2)\mu $eV~\cite{Birse:2012eb}, a proton radius value of 
$r_p= 0.84099(36)$~fm is obtained from $\mu$p~\cite{Antognini:2022xoo}.
This value is in agreement with the best and most recent determination from electron-nucleon
scattering and $e^+e^-$ annihilation data based on dispersion theory,
$r_p = 0.840^{+0.003}_{-0.002}{}^{+0.002}_{-0.002}$~fm~\cite{Lin:2021xrc}. These numbers agree
within errors, but clearly the muonic hydrogen result is more precise.
Note further that there has been (and still is) some tension with several other determinations from
H spectroscopy and electron-proton scattering, that continues to spark  lively discussions and
triggering more investigations across various fields, such as H spectroscopy
or further proton form factor measurements with electron and muon beams. 
For an update of the present situation  we refer to recent review articles, see e.g.
\cite{Lin:2021umz,Gao:2021sml,Peset:2021iul,Antognini:2022xoo,Karr:2020wgh}.

While the 2S-2P energy splitting is sensitive to electric properties of the proton as the
proton charge radius, the hyperfine splitting (HFS) is sensitive also to magnetic properties of
the proton as it arises from the interaction between the proton and muon magnetic moments.
To leading order, this interaction between magnetic moments yields an energy splitting
expressed in terms of the Fermi energy
\begin{equation}
E_\mathrm{F} = \frac{8 (Z\alpha)^4 m_r^3 (1+\kappa)}{3mM} =182.443~~\mbox{[meV]}~,
\end{equation}
where $m$ is the muon mass, $M$ the proton mass, $m_r$ the reduced mass of the $\mu$p system,
$\alpha$ the fine-structure constant and $\kappa$ the anomalous magnetic moment of the proton.
Radiative, recoil, relativistic and proton structure dependent contributions modify this energy splitting ~\cite{Eides:2000xc, Pachucki:1996zza, Antognini:2013rsa, Peset:2016wjq, Peset:2021iul}. 
For the HFS  of the ground state in $\mu$p the updated theory takes the form~\cite{Antognini:2022xoo}: 
\begin{equation}
  \begin{array}{cccccccc}
  E^\mathrm{th}_\mathrm{HFS}(\mu\text{p}) & =&  E_F     &+& \Delta E_\mathrm{QED} & + & \Delta E^{2\gamma} \label{eq:HFS-1}& \\
  & =&  182.443 &+& 1.354(7)                         & + & E_F\Big(1.01958(13)\Delta_\mathrm{Z}^{\mu{\rm p}}+ 1.01656(4)\Delta_\mathrm{recoil}^{\mu{\rm p}}+1.00402\Delta_\mathrm{pol}^{\mu{\rm p}}\Big)~~\mbox{[meV]} \: . %
  \end{array}
  \end{equation}
The second term $\Delta E_\mathrm{QED}=1.354(7)$~meV is the sum of all calculated QED contributions,
including minor weak  (Z-exchange) and hadronic vacuum polarization contributions.
For the HFS, the leading proton structure contribution is given by the two-photon-exchange
contribution $\Delta E^{2\gamma} $, which is conventionally divided into a Zemach radius
contribution $\Delta_Z^{\mu{\rm p}}$, a recoil contribution $\Delta_\mathrm{recoil}^{\mu{\rm p}}$ and a
polarizability contribution $\Delta_\mathrm{pol}^{\mu{\rm p}}$~\cite{Carlson:2008ke,Carlson:2011af,Tomalak:2017lxo,Tomalak:2017owk,Faustov:2006ve,Hagelstein:2015egb}.
While the sum of these three structure-dependent contributions is unambiguous,
the separation between the recoil and polarizability corrections depends upon a
protocol~\cite{Carlson:2008ke}, here  we use the formalism as presented in
Ref.~\cite{Hagelstein:2015egb}.

To give an idea of their sizes, these contributions are typically expressed in terms of the
Fermi energy $E_F$, and their value is about $\Delta_Z^{\mu{\rm p}} \approx 7500$~ppm, $\Delta_\mathrm{recoil}^{\mu{\rm p}}
\approx 850$~ppm and $\Delta_\mathrm{pol}^{\mu{\rm p}}\approx 350$~ppm~ (see e.g. \cite{Tomalak:2017lxo}).
The small deviations from unity of the numerical coefficients in Eq.~(\ref{eq:HFS-1}) arises from
radiative corrections. All three coefficients include wavefunction corrections caused by the
one-loop electron vacuum polarisation while the coefficient in front of $\Delta_Z^{\mu{\rm p}}$  accounts
also for the electron-vacuum polarisation insertion in the two-photon exchange diagram~\cite{Antognini:2022xoo}.

In a dispersive framework~\cite{Carlson:2008ke,Carlson:2011af,Tomalak:2017lxo,Tomalak:2017owk,Faustov:2006ve,Hagelstein:2015egb}, all the three contributions forming $\Delta E^{2\gamma} $ can
be expressed in terms of phenomenological (measurable) quantities of the proton structure.
The Zemach contribution $\Delta_Z^{\mu{\rm p}}$ that accounts for the elastic part of the two-photon exchange
contribution can be expressed through the electric, $G_E (Q^2)$, and  magnetic, $G_M(Q^2)$, Sachs form
factors: 
\begin{equation}
\label{DeltaZintegrand}
\Delta_Z  =-2Z\alpha m_r  \, r_\mathrm{Z}~,
%
\end{equation}
where $Z$ is the atomic number and 
$r_\mathrm{Z}$ the Zemach radius defined as~\cite{Zemach:1956zz}
\begin{equation}
  r_\mathrm{Z} = - \frac{4}{\pi} \int_0^\infty \frac{\mathrm{d} Q}{Q^2}
  \left[\frac{G_E(Q^2) G_M(Q^2) }{1+\kappa} -1\right]~.
  \label{Zemachradius}
\end{equation}
For early work on this moment of the charge/magnetization distribution of the proton, see e.g.
\cite{Zemach:1956zz,Friar:1978wv}, and for the most recent ones, see
e.g.~\cite{Lin:2021xrc, Borah:2020gte, Distler:2010zq}. For a precise definition of the
squared momentum transfer $Q^2$, see Sect.~\ref{sec:form}.

The so-called recoil contribution, which more precisely is the recoil correction to
the Zemach contribution, can also be described solely by form factors. In addition to
$G_E (Q^2)$ and  $G_M(Q^2)$ in this case also the Dirac  $F_1(Q^2)$ and Pauli $F_2(Q^2)$
form factors are used (see the Supplement of Ref.~\cite{Antognini:2022xoo}):
\begin{align}
  \Delta_\mathrm{recoil} &=\frac{Z \alpha}{\pi (1+\kappa)}\int_0^\infty \frac{\text{d} Q}{Q}
  \Bigg\{\frac{G_M(Q^2)}{Q^2}\frac{8m M}{v_l+ v }   \left(2F_1(Q^2)+\frac{F_1(Q^2)+3F_2(Q^2)}{(v_l+1)(v+1)}
  \right)\notag\\
  &\phantom{mmm}-\frac{8m_r G_M(Q^2)G_E(Q^2)}{Q}-\frac{m F_2^2(Q^2)}{M}\frac{5+4v_l}{(1+v_l)^2}\Bigg\}~,
  \label{Deltarecoilintegrand}
\end{align}
where $v= \sqrt{1+4M^2/Q^2}$ and  $v_l= \sqrt{1+4m^2/Q^2}$. For earlier calculations of
this quantity, see e.g. Refs.~\cite{Carlson:2008ke,Tomalak:2017lxo}. Recent work on the
recoil corrections can be found in Ref.~\cite{Pachucki:2022cuj}.
Differently the polarizability contribution that accounts for the inelastic part of the two-photon
exchange contribution  can be expressed through integrals over the inelastic structure functions
$g_i(x,Q^2)$ and the Pauli form factor $F_2(Q^2)$. The interested reader can find them e.g. in
Refs.~\cite{Carlson:2008ke,Carlson:2011af}.
Note that the polarisability contribution obtained from the dispersive
approach~\cite{Hagelstein:2015egb, Carlson:2008ke}  is
derived from the Compton scattering amplitude with finite proton mass so that in this framework
no recoil corrections to the polarizability contribution  are needed.

A precise evaluation of $\Delta_\mathrm{recoil}^{\mu{\rm p}}$  
is timely given the ongoing experimental efforts carried out by three collaborations
that aim at the HFS in $\mu$p~\cite{Amaro:2021goz, Kanda:2020mmc, Pizzolotto:2020fue} 
with relative accuracies ranging from 1 to 10~ppm.
While comparing with the measured HFS in muonic hydrogen, the theoretical prediction of
Eq.~(\ref{eq:HFS-1}) can be used to extract the total two-photon exchange contribution $\Delta E^{2\gamma} $,
the interpretation of the experimentally obtained  $\Delta E^{2\gamma} $ requires a precise
knowledge of the recoil contribution.
Indeed, in order to extract the polarisability contribution $\Delta_\mathrm{pol}^{\mu{\rm p}}$ or the
Zemach radius $r_\mathrm{Z}$ from the measured HFS, the recoil contribution $\Delta_\mathrm{recoil}^{\mu{\rm p}}$
has to be subtracted from the empirically determined $\Delta E^{2\gamma} $. The purpose of this
paper is thus to reduce the uncertainty of   $\Delta_\mathrm{recoil}^{\mu{\rm p}}$, presently on the
5~ppm level~\cite{Tomalak:2018uhr}, to maximize the physics interpretation of the HFS measurements
when they will be available.

From the theoretical side, the formalism can be straightforwardly
extracted from the muonic case to the H case, by replacing the muon
mass by the electron mass and correspondingly the reduced mass of the
lepton-proton bound state and the lepton velocity $v_l$.
For completeness and to give a sense of the size of the various corrections
 we report a summary of the theory in H in a
form analogous to Eq.~\eqref{eq:HFS-1}. The HFS for the ground state in H from
Ref.~\cite{Antognini:2022xoo} is
\begin{equation}
E^\mathrm{th}_\mathrm{HFS}(\text{H}) = 1418840.082(9) +1613.024(3)+E_F^{\rm H}
\Big(1.01558(13)\Delta_\mathrm{Z}^{\rm H}+ 0.99807(13)\Delta_\mathrm{recoil}^{\rm H}
+1.00002\Delta_\mathrm{pol}^{\rm H}\Big)~~\mbox{[kHz]}
\end{equation}  
where the  Fermi energy for hydrogen is $E_F^{\rm H}=1418840.082 (9)\,$kHz.

Evaluating $\Delta_\mathrm{recoil}^{\rm H}$  is interesting for the same reason as in $\mu$p,
i.e., for dissecting the polarizability and the Zemach radius
contributions from the measurement of the HFS in hydrogen.
Moreover, an improvement of $\Delta_\mathrm{recoil}^{\rm H}$
can also be used to improve on the prediction of two-photon exchange
contribution in $\mu$p via the scaling procedure presented in Ref.~\cite{Hagelstein:2015egb}.

This paper is organized in the following way.
Sect.~\ref{sec:form} contains a brief review of the underlying dispersion-theoretical
formalism and recalls the pertinent results from Ref.~\cite{Lin:2021xrc} used here.
The results for the recoil correction in muonic as well as electronic hydrogen are
displayed and discussed in Sect.~\ref{sec:summ}.

\section{Formalism}
\label{sec:form}

To set the stage, we briefly define the nucleon electromagnetic form factors. In fact, for the
dispersive analysis it is mandatory to consider protons and neutrons together,
for details see the review~\cite{Lin:2021umz}. Only later we will specialize to the
proton case (as already done in the introduction). These form factors are given
by the matrix element of the electromagnetic current $j^\mu$ sandwiched between
nucleon states,
\begin{equation}
\langle p' | j_\mu^{\rm em} | p \rangle = \bar{u}(p')
\left[ F_1 (t) \gamma_\mu +i\frac{F_2 (t)}{2 M} \sigma_{\mu\nu} q^\nu \right] u(p)\,,
\end{equation}
with $M$ the nucleon mass (either proton or neutron), $u(p)$ a conventional nucleon
spinor and $t=(p'-p)^2$  the four-momentum
transfer squared. In the space-like region of relevance here, one often
uses the variable $Q^2=-t>0$, cf. Eq.~\eqref{Deltarecoilintegrand}.
The form factors are normalized as
\begin{equation}
\label{norm}
F_1^p(0) = 1\,, \quad  \; F_1^n(0) = 0\,, \quad
F_2^p(0) =  \kappa_p\,, \quad  F_2^n(0) = \kappa_n\, ,
\end{equation}
with $\kappa_p=1.793$ and $\kappa_n=-1.913$ the anomalous magnetic moment of
the proton and the neutron, respectively. Also used are the Sachs form factors,
given by
\begin{equation}
G_{E}(t) = F_1(t) - \tau F_2(t) ~ , ~~ G_{M}(t) = F_1(t) + F_2(t) \, , 
\label{sachs}
\end{equation}
where $\tau = -t/(4 M^2)$. The proton charge radius $r_p$ follows as
\begin{equation}
\label{eq:rp}
 r_p^2 = 6 \frac{dG_p^E(t)}{dt}\Big|_{t=0}~.
\end{equation}

\begin{figure}[t!] 
\includegraphics*[width=0.6\linewidth,angle=0]{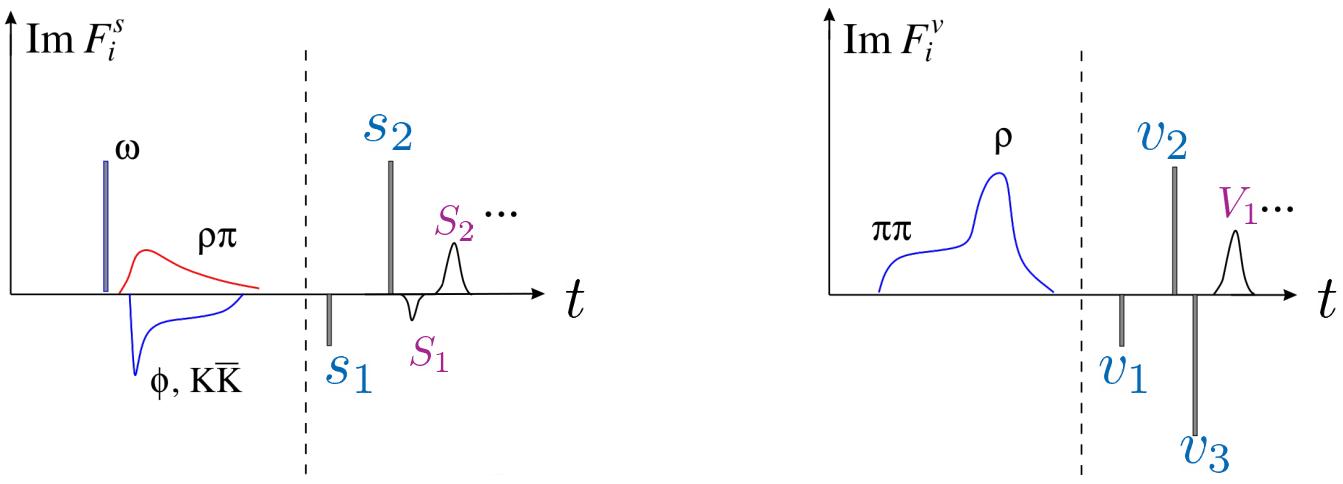}
\caption{Cartoon of the nucleon spectral function. Left panel: Isoscalar
  case. Here, the $\omega$ and $\phi$ mesons are relevant together with the
  $\pi\rho$ and $K\bar{K}$ continua, while $s_1, s_2, ...$ are narrow and $S_1, S_2, ...$ are
  broad effective poles. Right panel: Isovector case. Here, the $\pi\pi$ continuum not only
  generates the $\rho$ but is also visibly enhanced on the left shoulder of the $\rho$.
  Further,  $v_1, v_2, v_3, ...$ are narrow and $V_1, ...$ are  broad effective poles. 
}
\label{fig:spec}
\vspace{-3mm}
\end{figure} 

Next, we turn to the dispersive analysis of the nucleon electromagnetic form factors.
For a generic form factor $F(t)$, one writes down an unsubtracted dispersion
relation of the form:
\begin{equation}
F(t) = \frac{1}{\pi} \, \int_{t_0}^\infty \frac{{\rm Im}\, 
F(t')}{t'-t-i\epsilon}\, dt'\, ,
\label{eq:disp} 
\end{equation}
where $t_0$ is the threshold of the lowest cut of $F(t)$ and the $i\epsilon$ defines the
integral for values of $t$ on the cut. In fact, in the isospin basis, $t_0 = 4M_\pi^2$ in the
isovector and $t_0 = 9M_\pi^2$ in the isoscalar channel, respectively. The imaginary part ${\rm Im}\, F$,
the so-called spectral function, encodes the constraints from analyticity and unitarity besides
other important physics. These spectral functions are given in terms of continua, narrow vector
meson poles as well as broad vector mesons. In the isovector case, the spectral function can
be reconstructed up to about $\sim 1\,$GeV$^2$ from data on pion-nucleon scattering and
the pion vector form factor, as most precisely done in Ref.~\cite{Hoferichter:2016duk}. This
in fact not only generates the $\rho$-meson but also an important enhancement on the left
shoulder of the $\rho$, that is of utmost importance to properly describe the nucleon isovector radii.
In the isoscalar spectral function, the $\omega$-meson represents the lowest contribution, that is not
affected by uncorrelated three-pion exchange. Further up, in the region of the $\phi$-meson, there
is a strong competition between $K\bar{K}$ and $\pi\rho$ effects, which to some extent suppresses
this part of the spectral function. For momenta above $\sim 1\,$GeV$^2$, effective narrow poles
represent the physics at higher energies. To describe the observed oscillations
of the cross sections for $e^+e^-\to p\bar{p}$ and $e^+e^-\to n\bar{n}$  in the timelike region, 
additional broad poles are required. The spectral functions are further constrained by the
normalizations of the form factors given in Eq.~\eqref{norm} as well as the perturbative QCD behaviour,
$F_1 (t) \sim 1/t^2$ and $F_2(t) \sim 1/t^3$.  A cartoon of the spectral functions is
given in Fig.~\ref{fig:spec}.

The spectral functions are determined from a fit to the world data set on electron-proton scattering
as well as the reactions $e^+e^- \leftrightarrow \bar{p}p, \bar{n}n$, the latter giving the form factors
in the timelike region. The fit parameters are the vector meson masses (except for the $\omega$
and the $\phi$) and the residua as well as the widths for the broad poles. There are two sources
of uncertainties that need to be accounted for. First, the statistical error is obtained using
a bootstrap procedure and second, the systematic error is calculated from varying the number
of vector meson poles so that the total $\chi^2$ does not change by more than 1\%. A detailed
description of these methods is given in the review~\cite{Lin:2021umz}.

\begin{figure}[h!] 
\includegraphics*[width=0.45\linewidth,angle=0]{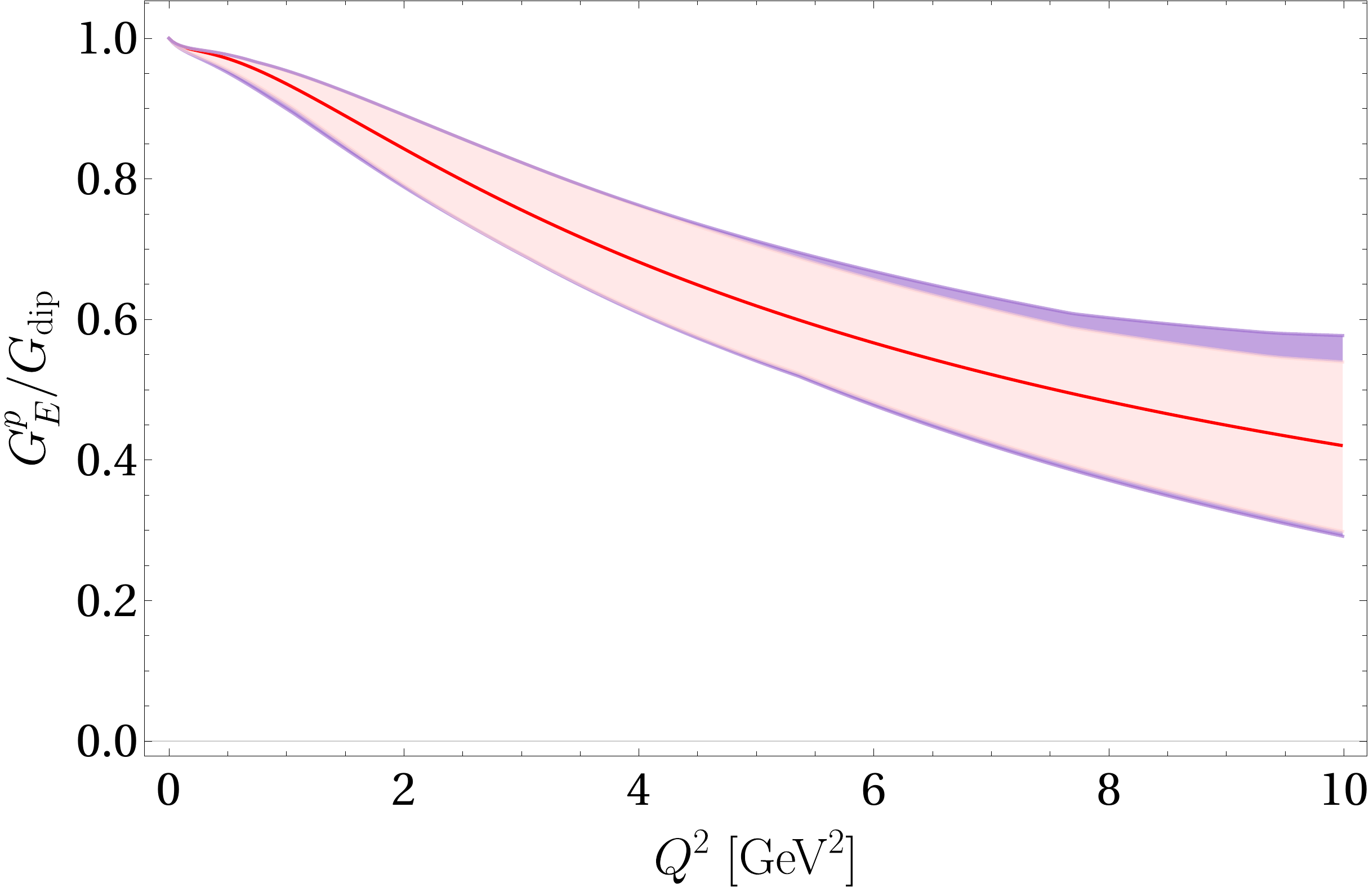}~~~~
\includegraphics*[width=0.45\linewidth,angle=0]{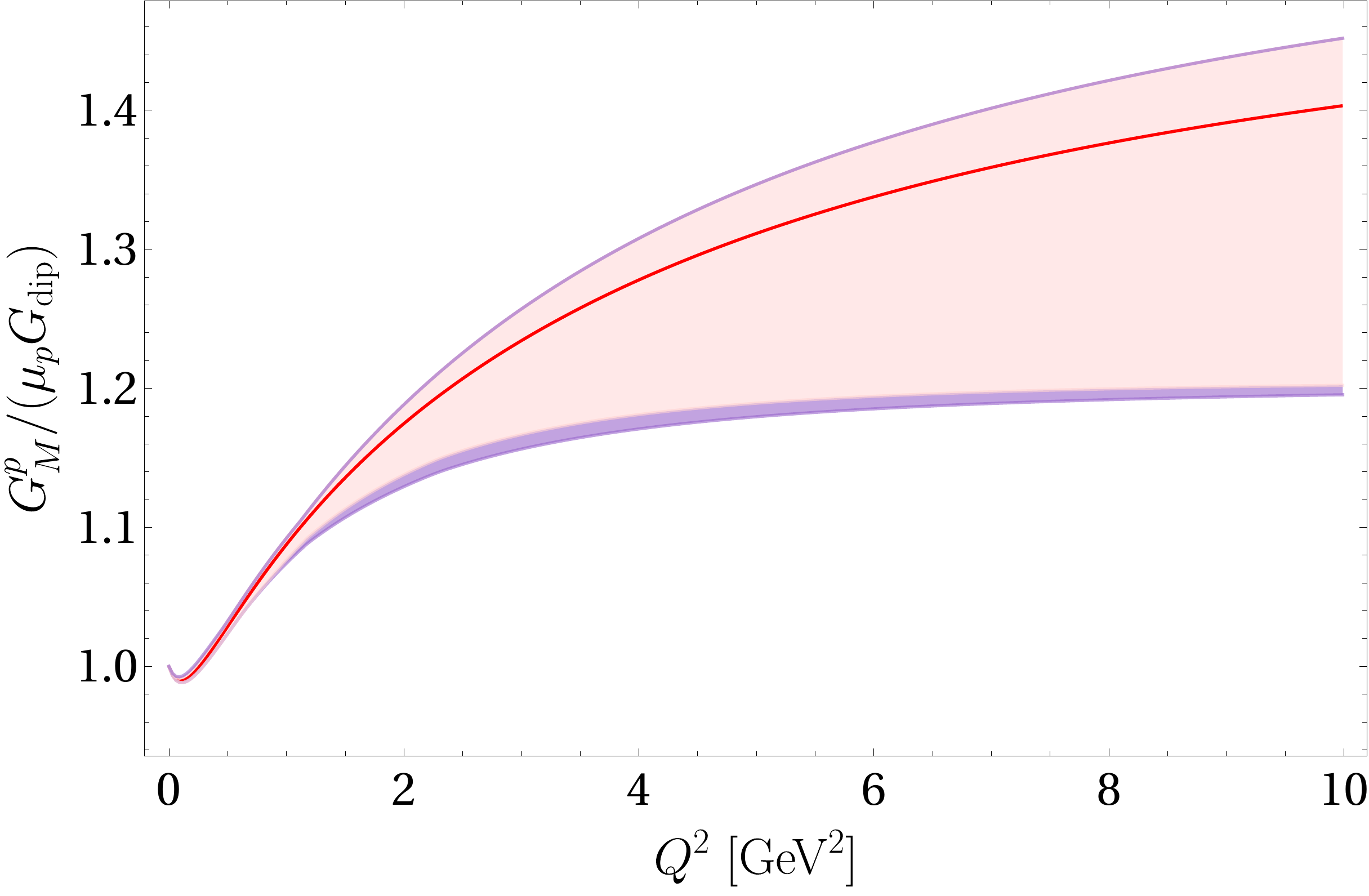} 
\caption{Electric (left panel) and magnetic (right panel) form factor of the proton
  from Ref.~\cite{Lin:2021xrc}
  divided by the canonical dipole form factor are shown by the red lines. The light red band is the
  statistical uncertainty and the purple band shows the systematic error added in quadrature. }
\label{fig:Gp}
\vspace{-3mm}
\end{figure}

The electric and magnetic form factors of the proton from Ref.~\cite{Lin:2021xrc}
normalized to the canonical dipole form, $G_{\rm dip}(Q^2) = (1+Q^2/0.71\,{\rm GeV}^2)^{-2}$, are shown in
Fig.~\ref{fig:Gp} together with their statistical and systematic uncertainties. From these,
the proton charge radius and the proton Zemach moment have already been extracted as~\cite{Lin:2021xrc}
\begin{equation}
  r_p = 0.840^{+0.003}_{-0.002}{}^{+0.002}_{-0.002}\,{\rm fm}~, \quad
  r_z = 1.054^{+0.003}_{-0.002}{}^{+0.000}_{-0.001}\,{\rm fm}~,
\end{equation}
where the first error is statistical and the second one is systematic.
These values are in good agreement with previous high-precision analyses
of the spacelike data alone \cite{Lin:2021umk,Lin:2021umz} and
have comparable errors.

\section{Results and Discussion}
\label{sec:summ}

We now turn to the calculation of the recoil correction defined in Eq.~\eqref{Deltarecoilintegrand}.
Consider first the $\mu$p system. We find
\begin{equation}
  \Delta_\mathrm{recoil}^{\mu{\rm p}} =  (837.6^{+1.7}_{-1.0}{}^{+2.2}_{-0.1}) \times 10^{-6}
  = (837.6^{+2.8}_{-1.0})  \times 10^{-6} = (837.6^{+2.8}_{-1.0})~{\rm ppm}~,
\end{equation}  
with the first error stemming from the bootstrap  and the last one from the variation
of the poles (systematic uncertainty). These errors are a few permile, so that this
can be considered as a high-precision determination. Compared with the most recent value
from Ref.~\cite{Tomalak:2017lxo}, $\Delta_\mathrm{recoil}^{\mu{\rm p}} = 844(5) \times 10^{-6}$,
these numbers agree  within errors but our result is more precise.

The analogous value for regular hydrogen is
\begin{equation}
  \Delta_\mathrm{recoil}^{\rm H} =   (526.9^{+1.1}_{-0.3}{}^{+1.3}_{-0.2}) \times 10^{-8}
  = (526.9^{+1.7}_{-0.4}) \times 10^{-8}~,
\end{equation}    
which is, as expected, two orders of magnitude smaller but with comparable
uncertainties as in the $\mu$p case.  Again, the corresponding number from
Ref.~\cite{Tomalak:2017lxo}, $\Delta_\mathrm{recoil}^{\rm H} = 532.8(4.9)\times 10^{-8}$,
is about 1\% larger but is also a bit less precise.\\\\

The Zemach radius can be extracted from the HFS measurement using the
theory of Eq.~(\ref{eq:HFS-1}), with $\Delta_\mathrm{recoil}^{\mu{\rm p}}$
from this study and assuming $\Delta_\mathrm{pol}^{\mu{\rm p}}$ from theory.
Similarly the polarizability contribution can be extracted using the
theory of Eq.~(\ref{eq:HFS-1}), with $\Delta_\mathrm{recoil}^{\mu{\rm p}}$
from this study and  taking the Zemach radius from e-p scattering or
from H spectroscopy.
Pinning down the uncertainty of this recoil--finite-size contribution allows
therefore to eliminate the most important higher-order
proton-structure dependent contribution that complicates and limits
extraction of the leading-order proton-structure effect (Zemach and
polarizability contributions) from the $\mu$p measurement.
The reduced uncertainty of $\Delta_\mathrm{recoil}^{\mu{\rm p}}$ from
this study can become particularly relevant in the scenario that the
smaller value of the polarizability contribution predicted by
the chiral perturbation theory will be confirmed.
Indeed there is presently an interesting tension between the
value of $\Delta_\mathrm{pol}^{\mu{\rm p}}$ predicted in a chiral
perturbation theory framework,  $\Delta_\mathrm{pol}^{\mu{\rm p}}=37(95)$~ppm
\cite{Antognini:2022xoo,Hagelstein:2018bdi, Hagelstein:2015lph},
and the values obtained from the data-driven approach, e.g. $\Delta_\mathrm{pol}^{\mu{\rm p}}=364(89)$~ppm
from Ref.~\cite{Tomalak:2017owk}.

Analogously, the reduced uncertainty of
$\Delta_\mathrm{recoil}^\text{H}$ can be used to improve on the
extraction of the polarizability contribution and the Zemach radius from the
HFS in H which has been measured with a
fractional accuracy of $7\times 10^{-13}$~\cite{Hellwig1970}.
The relative uncertainty of about $1\times 10^{-8}$ of $\Delta_\mathrm{recoil}^\text{H}$ 
set also the limit to which theory and experiment can be confronted
in H.
Testing the hydrogen HFS  beyond this relative accuracy requires
improving on the proton form factors.

Beside improving the interpretation of the $\mu$p HFS measurements
when they will be completed, the reduced uncertainty of
$\Delta_\mathrm{recoil}^\text{ H}$ can also be used to refine the
prediction of two-photon exchange contribution in $\mu$p using the scaling procedure
presented in Ref.~\cite{Hagelstein:2015egb}.
This serves to narrow down significantly the search range for the
HFS transition in $\mu$p easing considerably the ongoing experimental efforts.
%

\acknowledgments
This study has been initiated at the PREN2022 convention at Paris that was funded
within the EU Horizon 2020 research and innovation programme,
STRONG-2020 project under grant agreement No. 824093. We thank the organizers for
providing a very stimulating atmosphere.
UGM and YHL acknowledge the support of 
 the Deutsche Forschungsgemeinschaft (DFG, German Research
Foundation) and the NSFC through the funds provided to the Sino-German Collaborative  
Research Center TRR~110 ``Symmetries and the Emergence of Structure in QCD''
(DFG Project-ID 196253076 - TRR 110, NSFC Grant No. 12070131001),
by the Chinese Aca\-de\-my of Sciences (CAS) through a President's
International Fellowship Initiative (PIFI) (Grant No. 2018DM0034), and by the
VolkswagenStiftung (Grant No. 93562). 
AA acknowledges the support of the European Research Council (ERC) through CoG.
\#725039, and the Swiss National Science Foundation through the projects
SNF 200021\_165854 and SNF 200020\_197052.

\end{document}